\documentstyle[preprint,aps,pre]{revtex}
\begin{document}
\draft
\title{A New Phenomenology for the Disordered Mixed Phase}
\author{Gautam I. Menon\footnote{Email:menon@imsc.ernet.in}}
\address{The Institute of Mathematical Sciences, C.I.T. Campus, \\
Taramani, Chennai 600 113, India}
\date{\today}
\maketitle
\begin{abstract}

A universal phase diagram for type-II superconductors
with weak point pinning disorder is proposed. In this
phase diagram, two thermodynamic phase transitions
generically separate a ``Bragg glass'' from the
disordered liquid. Translational correlations in the
intervening ``multi-domain glass'' phase are argued to
exhibit a significant degree of short-range order.
This phase diagram differs significantly from the
currently accepted one but provides a more accurate
description of experimental data on high and low-T$_c$
materials, simulations and current theoretical
understanding.

\end{abstract}

Quenched randomness destabilizes the Abrikosov flux-line lattice in a
pure type-II superconductor, yielding novel glassy
phases\cite{review1}.  In the Bragg glass (BrG) phase\cite{giam},
translational correlations decay asymptotically as power laws.  On
increasing disorder or temperature, the BrG phase is unstable to a phase
in which such correlations decay exponentially\cite{giam}.  At large
values of the applied field $H$, the disordered liquid (DL) evolves
smoothly into a glassy state\cite{gammel} when the temperature $T$ is
reduced.  This glassy state may be a new thermodynamic phase, the
``vortex glass'', with spin-glass-type correlations in the
superconducting order parameter, separated from the disordered liquid
by a line of phase transitions\cite{fifi}.  An alternative picture,
attractive in many respects, is that this state resembles a structural
glass\cite{muon}, topologically disordered at the largest length scales
but lacking long-ranged phase correlations\cite{natterman1}.

Fig. 1 summarizes the currently popular view of phase behaviour in the
disordered mixed phase of type-II
superconductors\cite{giam,gammel,natterman1,popular,kiervin,khaykovich}.
The BrG phase is shown to be unstable to the vortex glass on increasing
$H$;  it also melts {\em directly} into the DL phase upon increasing
$T$.  The continuous vortex-glass to DL transition line meets the BrG-DL
transition line at a multicritical point\cite{khaykovich}.  Recent
studies\cite{gaifullin} indicate that the underlying field-driven Bragg
glass-vortex glass transition is first-order.

While this phase diagram {\em by construction} reflects early
experiments on YBCO\cite{gammel} and BSCCO\cite{khaykovich}, many
systems, among them the low-T$_c$ superconductors
2H-NbSe$_2$\cite{shobo,satya,mypaper},
CeRu$_2$\cite{mypaper,tenya,jsps} and
Ca$_3$Rh$_4$Sn$_{13}$\cite{shampa}, the cuprate NCCO\cite{giller1}, the
bismuthate (K,Ba)BiO$_3$\cite{klein}, the borocarbide
YNi$_2$B$_2$C\cite{jsps} and the mercury\cite{wisnewski} and
thallium\cite{hardy} based compounds, behave qualitatively
differently.  Even for the high-T$_c$ systems, more recent
experiments\cite{blasius} suggest features inconsistent with Fig. 1.
That these differences imply {\em qualitively} different types of phase
behaviour is one possibility.  Alternatively, such differences might
simply reflect different limits of an underlying universal phase
diagram; this possibility is both theoretically attractive and
economical. It would be sensible, however, to require that Fig. 1 be
recovered from such a phase diagram in the appropriate limit.

This Letter proposes Fig. 2 as a universal phase diagram for {\em all}
weakly disordered type-II superconductors.  The term ``multi-domain
glass'' (MG) describes the narrow sliver of glassy phase which we
propose {\em always} intervenes between BrG and DL phases.  This sliver
expands into the putative ``vortex glass'' phase\cite{fifi} of Fig. 1,
as $H$ is increased.  Experiments indicate that the MG phase melts via
a first-order phase transition on $T$ scans at intermediate $H$; this
first-order line is shown to meet a line of continuous transitions at a
tricritical point, although more complex topologies are possible.  The
inset shows the expected phase behaviour at $H \sim H_{c1}$.  The
predicted\cite{nelson} reentrance of the glassy phase at low fields was
established in Ref.\cite{ghosh}.  Recent experiments\cite{jsps,paltiel}
clearly show the disordered phase enveloping a quasi-lattice phase at
small $H$.  Fig. 2 incorporates recent proposals for such phase
behaviour in low-T$_c$ superconductors\cite{mypaper}, derived from an
analysis of peak effect phenomena in these systems\cite{jsps}, the
anomalous and sharp {\em increase} in critical currents j$_c$, seen
over a narrow regime below H$_{c2}$, the ``peak regime'', as $H$ or $T$
is varied.

The ``multi-domain'' glass is argued to be an {\em equilibrium} phase,
distinct from the vortex glass, consistent with recent
theoretical\cite{natterman1}, experimental and simulational
understanding.  Theoretically, screening destroys vortex-glass-type
order in three dimensions\cite{bokil}; recent
experiments\cite{strachan} question earlier reports of vortex-glass
scaling.  Simulations of {\em classical} directed lines interacting
with quenched point impurities see topologically disordered glassy
phases {\em in equilibrium} in a regime which intervenes between
relatively ordered (BrG) and disordered (DL) phases\cite{otterlo}.
Arguments for the equilibrium nature of this phase and a brief survey
of its glassy attributes in the context of low-T$_c$ materials were
provided in Ref.\cite{mypaper}.

What symmetry might be broken across the MG-DL
transition?  A recent study of disordered hard-sphere
fluids obtains, within mean-field theory, a one-step
replica symmetry breaking glass transition for
sufficiently strong disorder; these authors stress the
resemblance of the features of the mean-field phase
diagram they obtain to Fig.  1\cite{parisi}. Should
these ideas be applicable to the disordered mixed
phase, a {\em discontinuous} freezing of the
disordered density configuration characterizing the
glassy phase would be expected, as against relatively
{\em smooth} behaviour of the entropy and internal
energy across this transition.  An alternative
possibility for the MG phase, consistent with some of
the proposals here and testable via an analysis of
Bitter decoration patterns, is a hexatic
glass\cite{hexatic}; the recent discovery\cite{dna} of
an equilibrium hexatic phase in a dense system of long
oriented DNA molecules favours a similar possibility
for line vortices.

We argue that translational correlations in the MG
phase can be fairly long-ranged (unlike correlations
in typical glassy phases), and propose that structure
in the MG phase just above the BrG-MG phase boundary
is best described as a mosaic of ordered domains, with
typical scale R$_d \gg a$, the inter-line spacing, for
weak pinning.  R$_d$ should be {\em largest} in the
interaction-dominated regime but should {\em decrease}
rapidly for much larger or smaller $H$, reflecting the
increased importance of disorder both at high-field
and low-field ends\cite{mypaper,jsps,ghosh,note}.
Some diagnostics for the multi-domain character of
this phase are the following:  The transition between
BrG and DL phases can occur via several intermediate
stages, as individual domains melt as a consequence of
disorder-induced inhomogeneities in the melting
temperature\cite{soibel}.  Fluctuations of a
domain-like arrangement can yield noise signals of
large amplitude associated with relatively {\em few}
fluctuators\cite{shobo,merithew}, intermediate
structure in the ac susceptibility as a function of
$T$\cite{mypaper,jsps}, a stepwise expulsion of
vortices on cooling\cite{ooi1} and a host of related
features\cite{mypaper,menon} of the experimental data
which are hard to rationalize in other pictures.

The MG-DL transition is interpreted as the true
remnant of the underlying freezing transition in the
pure system. The presence of a critical point on this
transition line\cite{kiervin} follows from the
observation that signals of melting should cease to be
seen once $R_d$ is comparable to correlation lengths
in the pure liquid at freezing.  This phenomenology
also rationalizes\cite{mypaper} a large body of data
on anomalies associated with the peak effect
phenomenon\cite{shobo}, including ``fracturing'' of
the flux-line array\cite{satya}, the association of
thermodynamic melting with the transition into the
disordered liquid\cite{ishida} as well as conjectures
regarding the dynamic coexistence of ordered and
disordered phases in transport measurements in the
peak regime\cite{paltiel}.

A simple conjecture for the scale of such domain sizes suggests
R$_d \sim R_a$, a Larkin length, at least at intermediate $H$.
This length is computed using the coarse-grained
free energy\cite{review1}
$F_{el} = \int d{\bf r}_\perp dz~[\frac{c_{11}}{2}
({\bf \nabla}_\perp \cdot {\bf u})^2 
+ \frac{c_{66}}{2} ({\bf \nabla}_\perp \times {\bf u})^2 + 
\frac{c_{44}}{2} (\partial_z {\bf u})^2 ]$,
where ${\bf u}({\bf r}_\perp,z)$ is the displacement field 
at location (${\bf r}_\perp,z$) and the
integral represents the elastic
cost of distortions from the ideal crystalline state with vortex
lines centred at ${\bf R}_i$, governed
by the values of the elastic constants for shear (c$_{66}$),
bulk (c$_{11}$) and tilt (c$_{44}$)\cite{review1}.
Quenched random pinning is incorporated by adding
$F_{d} = \int d{\bf r}_\perp dz~V_d({\bf r_\perp},z)\rho({\bf r}_\perp,z)$,
to F$_{el}$,
where $\rho({\bf r}_\perp,z) = \sum_i \delta^{(2)}
({\bf r}_\perp - {\bf R}_i - {\bf u}({\bf R}_i,z))$. This term models
the interaction with a quenched Gaussian disorder potential $V_d$, 
assumed to be correlated over $\xi$, the coherence length {\it i.e.}
$[V_d(x)V_d(x^\prime)] = K(x - x^\prime)$, with $K(x-x^\prime)$ a
function of range $\xi$\cite{natterman1}. The notation $x$
denotes (${\bf r_\perp},z$).

The correlator $B({\bf r_\perp},z) = [\langle {\bf u}({\bf r}_\perp,z)
- {\bf u}({\bf 0},0) \rangle^2]$ where $\langle\cdot\rangle$ and
$[\cdot]$ denote thermal and disorder averages respectively, has the
following properties\cite{natterman1}:  For $r_\perp,z \ll R_c,L^b_c$
(the transverse and longitudinal Larkin pinning lengths), correlations
behave as in the Larkin ``random force'' model {\it i.e.} $B(x) \sim
x^{4-d}$. At length scales between $R_c$ and $R_a$, the Larkin length
scale referred to above at which disorder and thermal
fluctuation-induced positional fluctuations become of order $a$, $B(x)
\sim x^{2\zeta_{rm}}$ with
$\zeta_{rm} \sim (4-d)/6 \sim 1/6$. The Larkin
length scale can then be estimated via $R_a \simeq
R_c(\frac{a}{\xi})^{\large {1/\zeta_{rm}}}$.  At still larger length
scales, $B(x) \sim$ log(x)\cite{natterman1}.

We estimate the Larkin pinning lengths R$_c$ and $L^b_c$ from $L^b_c =
\frac{2\xi^2c_{66}c_{44}}{nf^2}$, $R_c =
\frac{\sqrt{2}\xi^2c^{3/2}_{66}c^{1/2}_{44}}{nf^2}$, and V$_c \sim
R_c^2L_c$. Equating $BjV_c$ to the energy gain from random pinning
$f(nV_c)^{1/2}$ yields the standard weak-pinning expression $j_c =
\frac{1}{B}f\left(\frac{n}{V_c}\right)^{1/2}$. If $R_c, L^b_c >
\lambda$, $j_c \sim
\left(\frac{\xi}{R_c}\right)^2j_{dp}$ and $L^b_c \sim \frac{\lambda}{a}
R_c$\cite{review1}, with j$_{dp}$ the depairing current.  For 
2H-NbSe$_2$: T$_c \simeq 7 K$, $\lambda \simeq 700 \AA$ ($H \parallel
c$), $\xi \sim 70\AA$ and $a \sim 450 \AA $ at $B \sim 1T$. Using
$j_c/j_{dp} \sim  10^{-4}$, we obtain transverse and longitudinal
Larkin {\it pinning} lengths $\frac{R_c}{a}\sim 15$ and
$\frac{L^b_c}{a}\sim 24$.

At fields $\sim 1T$, $(\frac{a}{\xi})^{1/\zeta_{rm}} \sim
(\frac{450}{70})^6 \sim 7\times 10^4$ leading to $R_d/a \sim R_a/a \leq
10^6$. Assuming, conservatively, $R_d
\sim 10^4$a and a longitudinal Larkin length comparable to the size of
the sample in the c-axis direction, an estimate for the number of
``independent fluctuators'' obtained in noise measurements on
2H-NbSe$_2$\cite{merithew,satya} can be obtained. For a sample in the
shape of a thin platelet of transverse area $A \sim 1 mm \times 1 mm$
and with $A_d \sim R_d^2$, the number of such fluctuators $N_f \sim
\frac{A}{A_d} \sim 10^1-10^2$, numbers small enough to yield strong
non-Gaussian effects\cite{merithew}.

Experimentally, behaviour in the mixed phase of low-T$_c$ materials
such as 2H-NbSe$_2$, CeRu$_2$, Ca$_3$Rh$_4$Sn$_{13}$, YNi$_2$B$_2$C and
several related compounds exhibit a remarkable
commonality\cite{jsps,mypaper}.  In these relatively pure materials
(j$_c$/j$_{dp} \sim 10^{-4}$) where thermal fluctuations are
substantial (Ginzburg numbers G$_i \sim 10^{-4}$ here, as against
typical values for low-T$_c$ systems of about $10^{-8}$), there is now
considerable evidence for a two-step transition from a relatively
ordered low-$T$ phase into a highly disordered fluid
phase\cite{jsps,mypaper}.  This transition occurs via an intermediate
and profoundly anomalous regime with glassy properties\cite{mypaper}.
These experiments indicate that the two transition lines which separate
the BrG from the DL phase can always be separately resolved.  The data
on NCCO\cite{giller1} and (K,Ba)BiO$_3$\cite{klein} are consistent with
Fig. 2 but not with Fig.  1.

Hall-probe based magnetization measurements on BSCCO favour a
single-step transition into the disordered liquid out of the BrG
phase\cite{zeldov}.  However, recent muon-spin rotation experiments see
{\em two} transitions\cite{blasius}.  The asymmetry of the field
distribution, $\alpha = \langle\Delta B^3\rangle^{1/3}/\langle\Delta
B^2\rangle^{1/2}$, exhibits a first transition in which $\alpha$ jumps
from a value of about 1.2 (consistent with a vortex-line crystal) to a
value of about 1 (indicating a considerable degree of local
translational order) as $T$ is increased. Across a second transition
boundary, obtained on further increasing $T$, $\alpha$ drops abruptly
to values close to zero, characteristic of the fluid.

Early experiments on YBCO found that melting in relatively clean
untwinned crystals was best described as a ``complex, two-stage
phenomenon''\cite{danna1}.  Recent simultaneous measurements of ac
susceptibility and magnetization in this material\cite{ishida} see a
sharp peak effect in which the location of the peak in ac
susceptibility correlates exactly to the magnetization jump which
signals melting.  The regime in which $\chi '$ shows non-trivial
signatures of the transition is enlarged, compared to the magnetization
jump which occurs at a sharply defined temperature, a feature which
follows transparently from the discussion here -- the width of the
transition region is related to the width of the sliver phase, while
the most prominent signatures of melting should generically be obtained
across a {\em line}, the MG-DL transition line.

When does the phase diagram of Fig. 1 approximate that
of Fig 2?  The sliver of MG phase is exceedingly
narrow at intermediate $H$ and for weak disorder, a
consequence of the reduction of effective disorder in
an interaction-dominated regime\cite{mypaper}.
Thermal smoothening of disorder should reduce the
width of this sliver further; thermal fluctuations
smear the disorder potential $U_p(T)$ seen by a vortex
line, strongly renormalizing the {\em effective}
disorder if $[<u^2>] \geq \xi^2$.  For a single line
pinned by quenched disorder $U_p(T) \sim
U_p(0)\exp[-c(T/T_{dp})^3]$\cite{popular,nelson} where
U$_p(0)$ is the pinning potential per unit length at
$T=0$, $c$ is a numerical constant and T$_{dp}$ is the
depinning temperature; $T_{dp} \sim
(U_p^2\xi^3\epsilon_0/\gamma^2)^{1/3}$ where
$\epsilon_0 = (\Phi_0/4\pi\lambda)^2$, and $\gamma$ is
the mass anisotropy. For YBCO, T$_{dp} \sim
20-30K$\cite{ssten}.  Such a substantial (exponential
in the simple limit above) reduction in effective
disorder should render the sliver unobservable in many
types of experiments, yielding an {\em apparent}
single melting transition out of the ordered phase.

The ratio $R = T^{low}_{dp}T^{high}_m/T_{dp}^{high}T_m^{low}$ measures
the relative importance of thermal fluctuations in low and high T$_c$
materials; T$_m$ is the melting temperature in the pure case and we
have approximated T$_m \sim $T$_c$.  Using the following values:
$NbSe_2:  U_p = 10K/\AA, \xi = 70\AA, \lambda = 700\AA, \gamma = 5$ and
$YBCO: U_p = 10K/\AA, \xi = 20\AA, \lambda = 1400\AA, \gamma = 50$ we
obtain $R \sim 10^3$, suggestive of the relative importance of this
effect to the high-T$_c$ materials {\it vis. a vis.} its irrelevance in
low-T$_c$ systems except very close to H$_{c2}$.  Experimentally, the
nominal irreversibility line for $H \sim 3-7T$ in YBCO lies well {\em
below} the melting line in weakly disordered samples\cite{ssten}. Thus,
thermal melting occurs in a nearly reversible regime.  Local Hall
probe-based susceptibility measurements on YBCO \cite{billon} find that
j$_c$ is actually finite below T$_m$ but extremely small ($\sim 0.4
A/cm^2$). These experiments see a peak effect close to and {\em below}
T$_m(H)$, a feature {\em inaccessible} in usual SQUID based
magnetization experiments. Other experiments at somewhat smaller $H$
see a very clear and sharp peak effect\cite{ishida}, with properties
similar to those in the low-T$_c$ materials\cite{mypaper}. These
results illustrate that signals of a two-step transition in high-T$_c$
materials may be very hard to access, particularly if discontinuities
in $j_c$ or magnetization across the first transition are small.  In
contrast, experiments on the low-T$_c$ materials discussed in
Refs.\cite{jsps,mypaper} indicate an irreversibility line located in
the {\em fluid} phase and provide clear evidence of a two-step
transition.

Anomalously {\em small} values of j$_c$ translate to anomalously {\em
large} values of $R_d$.  The magnetization discontinuity across the
BrG-MG phase boundary should scale roughly as the density of unbound
dislocations $\rho_d$ in the MG phase for small $\rho_d \sim 1/R_a^2$,
yielding $\Delta M \sim \Delta M_0 (a/R_a)^2$ where $\Delta M_0$ is the
magnetization jump in the pure system at T$_m$. Even if R$_a/a \sim
30$, the corresponding induction jump $\Delta B \sim \Delta M$ is of
order $10^{-3}M_0$ or within noise levels in a typical
experiment\cite{zeldov,soibel}.  Thus, for sufficiently weak disorder,
the sample may behave substantially as a single domain in the
intervening MG phase, implying the absence of signals of the BrG-MG
transition in a magnetization experiment and thus the phenomenology of
Fig. 1.

The apparent vanishing of the sliver rationalizes the putative
multicritical point\cite{gammel,popular} of Fig. 1.  While the sliver
may not be resolvable in some classes of experiments, it may be
apparent in others, particularly those which probe local correlations.
However, in the thermodynamic limit, provided other, unrelated direct
instabilities to the liquid do not intervene, continuity suggests two
transitions as in Fig.  2. In essence, the phenomenology of Fig. 1 {\it
assumes} by {\it fiat} such an instability.  There appears to be little
justification for this assumption; we suggest that it need not hold.
For more disordered superconductors, whether high-T$_c$ or low-T$_c$, a
two-step transition should generically be seen.

Many recent simulations, for example Ref.\cite{otterlo}, see an
intermediate field MG phase with translational correlations comparable
to system sizes, although the dynamics changes abruptly across the
BrG-MG transition.  Interestingly, Ref.\cite{otterlo} concludes that the
existence of a sliver of MG phase always preempting a direct BrG-DL
transition cannot be ruled out, as in the recent simulations of 
Sugano {\it et.  al.} \cite{sugano}.

In conclusion, it is suggested that the BrG phase in all disordered
type-II superconductors {\em generically} transforms first into an
intermediate glassy state on heating, rather than directly into a
liquid; the proposal of a generic two-step transition out of the
disordered liquid state should be experimentally testable in
thermodynamic measurements.  More details will appear
elsewhere\cite{menon}.

I thank C.  Dasgupta for a critical reading of the manuscript.  These
ideas evolved out of many discussions with S. Bhattacharya; his
insights were crucial to this work.  D.R.  Nelson is thanked for a
useful conversation regarding hexatic glasses.  Support from a DST
(India) Fast Track Fellowship for Young Scientists is gratefully
acknowledged.

\begin{figure}
\caption{The current view of the phase diagram of disordered type-II
superconductors\protect\cite{gammel,natterman1,popular,kiervin}.
In addition to Meissner and normal phases, this
phase diagram subdivides the mixed phase into three phases -- the 
Bragg glass (BrG), the vortex glass (VG) and the disordered 
liquid (DL); these are described in the text. Note that the
Bragg Glass phase melts directly into the liquid
at intermediate field values.
}
\end{figure}
\begin{figure}
\caption{Proposed universal phase diagram for disordered type-II
superconductors. The Meissner, normal, BrG and DL phases are as 
shown in Fig. 1. The MG phase (see text) intrudes between
BrG and DL phases everywhere in the phase diagram.  The inset expands 
the boxed region of the main figure, illustrating the reentrance 
of the BrG-MG phase boundary. Solid lines indicate discontinuous 
transitions.
}
\end{figure}


\begin{thebibliography}{99}
\bibitem{note} The increased importance of disorder at
low fields should imply a weakening of the first-order
behaviour of the MG-DL transition line of Fig. 2 close
to H$_{c1}$, mirroring the behaviour at high-fields.
This would suggest another tricritical point at the
low-field end of the MG-DL transition line, as seen in
some recent experiments.
\end{thebibliography}
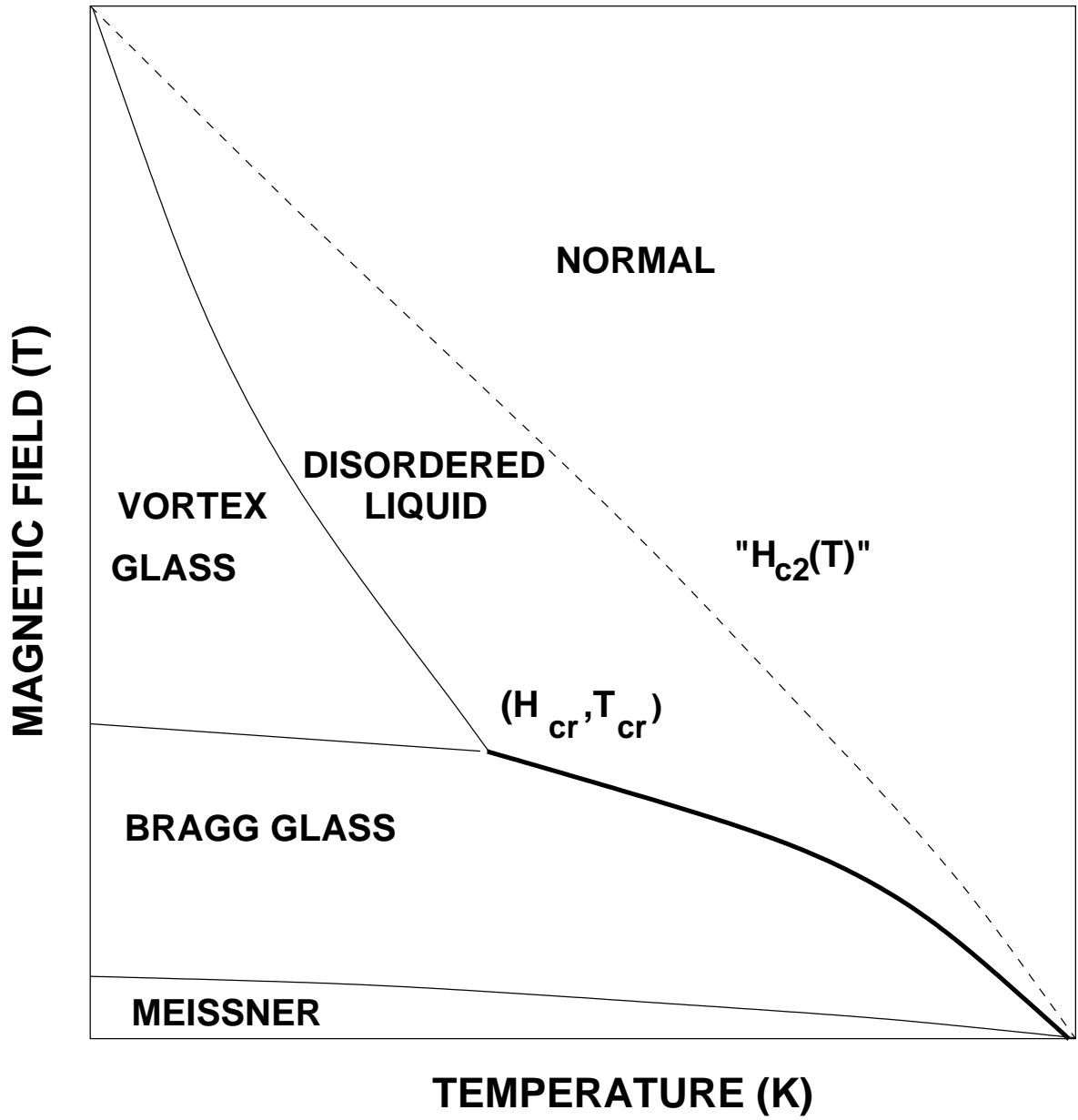
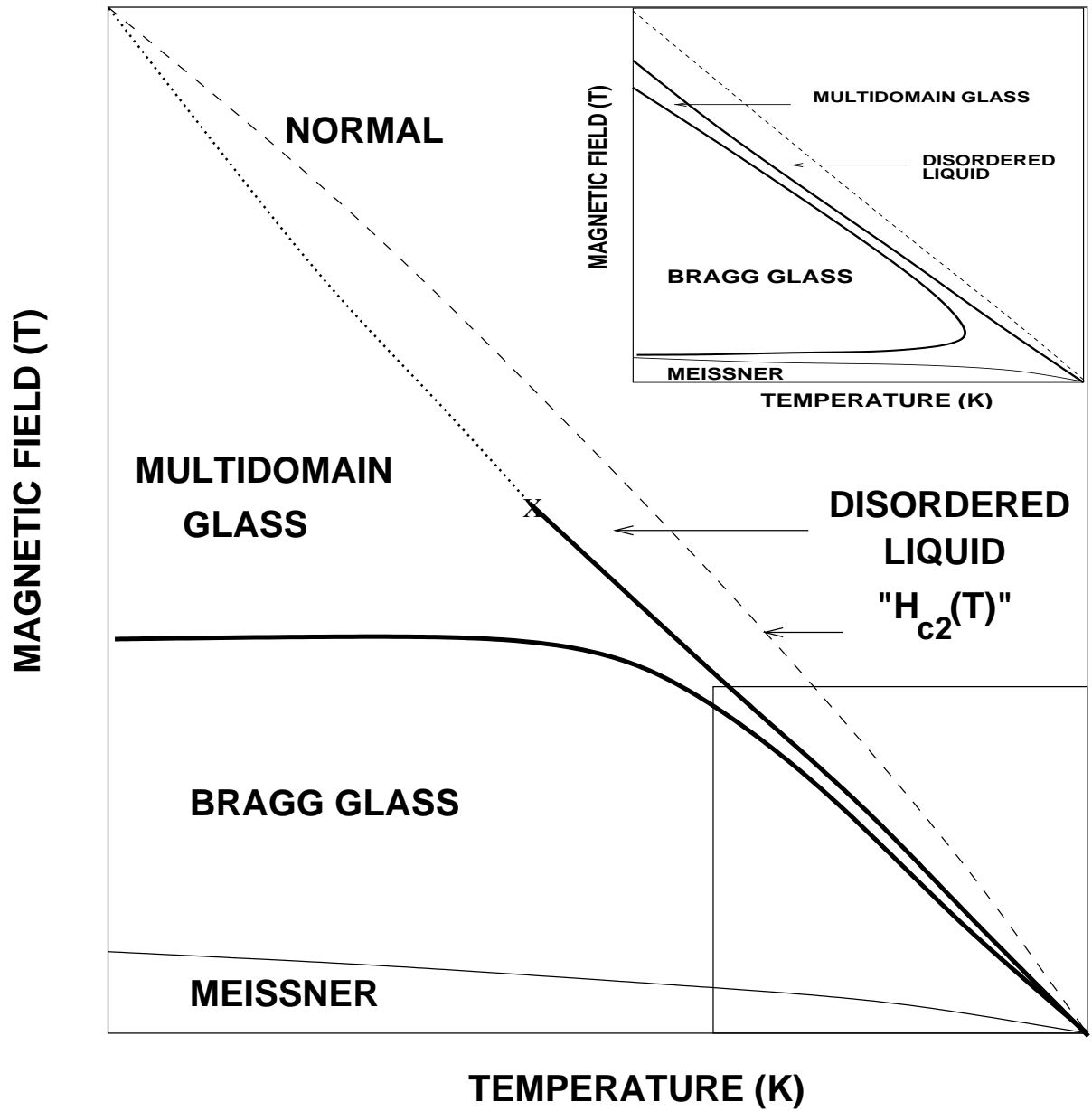
\end{document}